\documentclass[%showpacs, 
twocolumn, preprintnumbers, amsmath,amssymb,nofootinbib]{revtex4}
\usepackage{graphicx}% Include figure files
\usepackage{dcolumn}% Align table columns on decimal point
\usepackage{bm}% bold math

\newcommand\ee{\end{equation}}
\newcommand\be{\begin{equation}}
\newcommand\eea{\end{eqnarray}}
\newcommand\bea{\begin{eqnarray}}
\newcommand{\sfrac}[2]{{\textstyle\frac{#1}{#2}}}
\newcommand\di{\partial}
\newcommand\mpl{M_{\rm Pl}}

\def\vx{{\vec x}}
\def\vk{{\vec k}}
\def\vq{{\vec q}}

\begin{document}

%\preprint{}

\title{One-particle-irreducible consistency relations for cosmological perturbations}% Force line breaks with \\

\author{Walter D. Goldberger}
\email{walter.goldberger@yale.edu}

\affiliation{
Physics Department, Yale University, New Haven, CT 06520, USA
}

\author{Lam Hui}
\email{lhui@astro.columbia.edu}

\author{Alberto Nicolis}
\email{nicolis@phys.columbia.edu}

\affiliation{%
Physics Department and Institute for Strings, Cosmology and Astroparticle Physics,\\
Columbia University, New York, NY 10027, USA
}%

\date{\today}% It is always \today, today,

\begin{abstract}
We derive consistency relations for correlators of scalar cosmological perturbations which hold in the ``squeezed limit" in which one or more of the external momenta become soft.   Our results are formulated as  relations between suitably defined one-particle irreducible $N$-point and $(N-1)$-point functions that follow from residual spatial conformal diffeomorphisms of the unitary gauge Lagrangian.   As such, 
some of these  relations are \emph{exact} to all orders in perturbation theory, and do not rely on approximate deSitter invariance or other dynamical assumptions (e.g., properties of the operator product expansion or the behavior of modes at horizon crossing).    The consistency relations apply model-independently to cosmological scenarios where the time evolution is driven by a single scalar field.  Besides reproducing the known results for single-field inflation in the slow roll limit, we verify that our consistency relations hold more generally, for instance in ghost condensate models
{\em in flat space}.
%, even in a limit in which the background metric becomes Minkowski.     
We comment on possible extensions of our results to multi-field models.

\end{abstract}

%\pacs{\dots}% PACS, the Physics and Astronomy
\maketitle

\vspace{.3cm}

\section{Introduction}

While the observed temperature anisotropies of the CMB are consistent with the possibility that the early universe underwent a period of exponential inflation, there is as of yet no definitive test of the dynamics of  inflation.    In the near future, experiments such as Planck will provide further constraints from studies of the CMB $B$-modes, which would probe the spectrum of gravitational waves produced during inflation, and possibly of non-Gaussian correlations of the temperature anisotropies.

The possibility of observing non-Gaussian features in the CMB and large scale structure 
has motivated much theoretical work in the last decade aimed at understanding the predictions of inflation for three- and higher point correlation functions of  density perturbations.    In the context of single field, slow-roll inflation a systematic treatment of scalar and tensor non-Gaussianity was first given in~\cite{Maldacena:2002vr}.   The explicit results of ref.~\cite{Maldacena:2002vr} indicate the existence of a set of \emph{consistency relations} between the two-point and three-point correlation functions, which hold in the soft  (or ``squeezed") limit in which one of the momenta approaches zero.    For the usual mode $\zeta$ that describes adiabatic density perturbations, this relation takes  the form 
\begin{equation}
\label{eq:1}
\langle \zeta_{\vk\rightarrow 0} \zeta_{\vk_1} \zeta_{\vk_2} \rangle = - (n_s -1) \delta({\vk}_1+{\vk}_2) P({\vk}\rightarrow 0) P({\vk}_1)
\end{equation}
where $P({\vk})\sim k^{-3+(n_s-1)}$ is the power spectrum of adiabatic modes.   Heuristically, this relation can be understood to follow from the behavior of long wavelength adiabatic modes~\cite{Maldacena:2002vr}.   Such modes freeze out at horizon crossing, becoming indistinguishable from a re-definition of the background scale factor.   Therefore, an equal-time correlator with one insertion of a long wavelength $\zeta$ is equivalent to a correlator for the remaining fields, evaluated at re-scaled coordinates.   This line of reasoning was subsequently used in~\cite{Creminelli:2004yq} to argue that the relation Eq.~(\ref{eq:1}) is in fact a general statement about single field inflation, independent of the slow roll approximation.

 Since the work of~\cite{Maldacena:2002vr,Creminelli:2004yq}, much recent effort has been dedicated to further understanding and generalizing the original consistency relations such as Eq.~(\ref{eq:1}).  Ref.~\cite{Senatore:2012wy} considered the behavior of $N$-point correlators in the limit of soft \emph{internal} momenta in which linear combinations approach zero.   New single-field relations relating  gradients of $(N+1)$-point functions and $N$-point correlators have been derived in~\cite{Creminelli:2012ed}.   These results were obtained both on the basis of approximate deSitter invariance (in the spectator field limit) and more generally, at tree level but all orders in slow roll, by
exploiting a relation~\cite{Weinberg:2003sw} between long wavelength adiabatic modes and ``large" spatial diffeomorphisms,  i.e. dilations and special conformal transformations, acting on a homogeneous background.  See~\cite{Hinterbichler:2012nm} for a generalization of this approach.

 These large diffeomorphisms also play a role in the work of~\cite{Assassi:2012zq}, which approaches the dilation consistency relations from the point of view of Ward identities associated with non-linearly realized dilation invariance, with $\zeta$ playing the role of the Goldstone mode.    In addition to contributions from single $\zeta$ poles, which reproduce~\cite{Creminelli:2012ed}, ref.~\cite{Assassi:2012zq}, finds additional contributions from higher Fock states,  although these are found to be subleading for single field models with Bunch-Davies initial vacuum state.  Ref.~\cite{Assassi:2012et} proves that superhorizon constancy of $\zeta$ holds as an operator statement in single-field scenarios with Bunch-Davies initial conditions, closing a possible loophole in the derivation of~\cite{Creminelli:2012ed}.    Relations between correlators in multi-field models, obtained via approximate deSitter invariance and the operator product expansion, have been obtained in~\cite{Kehagias:2012pd}, while~\cite{Schalm:2012pi} uses approximate $SO(4,1)$ invariance of the background to reproduce the relation between two- and three-point functions.

It should be pointed out, however, that despite these recent developments, there seems to be no universal consensus on the status of the consistency relations, even in single-field theories.   A proposed counterexample was given in~\cite{Chen:2013aj}, while the analysis presented in~\cite{Agarwal:2012mq} indicates the relation in Eq.~(\ref{eq:1}) is violated for generic initial states.  Ref.~\cite{Tanaka:2011aj} argues that although  Eq.~(\ref{eq:1}) is correct, it does not reflect measurements made by physical observers.   Rather, physical (gauge invariant) non-Gaussianities should have softer behavior in the squeezed limit than what is suggested by Eq.~(\ref{eq:1}).

 In this paper, we present consistency relations that follows from residual diffeomorphism invariance of ``$\zeta$-gauge"  scalar cosmological perturbations.    The main assumptions that go into our derivation are gauge invariance of  the action and of the path integral measure.   Because the set of starting assumptions are relatively small,  our results are expected to be quantum mechanically exact, and to have a wider scope of validity than just single field, slow-roll scenarios.  In particular, our relations hold independently of dynamical assumptions (approximate isometries, superhorizon behavior, etc.), or of the details of the initial state.   We find it most natural to express our results in terms of in-in, one-particle irreducible (1PI) correlators of $\zeta$ at equal times, rather than the more conventional connected Green's functions used  to describe non-Gaussianity.  However, besides implying new, non-trivial, squeezed relations between the two- and three-point function, our 1PI  relations are completely consistent with those obtained in the literature under more restrictive assumptions.

In sec.~\ref{formalism} we introduce the effective action $\Gamma[\zeta]$ that generates in-in 1PI correlators of $\zeta$ at equal times, and derive formal constraints due to invariance under spatial diffeomorphisms.
Explicit results for the special cases of residual spatial dilation and special conformal invariance are given in sec.~\ref{formalism}A and sec.~\ref{formalism}B respectively.   These consist of relations between $N$- and $(N+1)$-point correlators with one zero-momentum mode (sec.~\ref{formalism}A) and between derivatives at zero momentum (sec.~\ref{formalism}B).   When re-expressed in terms of connected correlators, our dilation relations agree with the standard results in the literature.   For the case of special conformal transformations, which do not preserve the closure of the momentum space polygon, we find that a procedure for computing derivatives of correlators must be specified in order to obtain well defined relations.   For any such procedure, our results are similar in form to those of~\cite{Creminelli:2012ed} and also of~\cite{Maldacena:2011nz} involving pure deSitter space correlators. 

Explicit examples are give in sec.~\ref{sec:examples}.   Sec.~\ref{sec:exsr} applies our relations to slow-roll inflation.    In particular, by working in terms of 1PI correlators, we are able to formulate and verify a new special conformal relation between the two-point and three-point functions.   In sec.~\ref{ghost}, we check that the consistency relations also hold for scalar perturbations of the ghost condensate~\cite{ArkaniHamed:2003uy}, for a choice of model parameters in which the background spacetime is tuned to be exactly Minkowski.    This example serves to illustrate that the consistency relations do not require the presence of a nearly-deSitter cosmological horizon and the freezing out of long wavelength modes in order to hold.

However, we stress that our results are not completely universal.  In sec.~\ref{counter} we present a simple model that \emph{violates} the consistency relations, namely a free massless scalar field in flat spacetime, decoupled from gravity.   This theory admits a spatially homogeneous
 background solution $\phi \propto t$ with constant energy density and equation of state $p=\rho$.  It has in addition adiabatic perturbations $\zeta$ that start at quadratic order in the underlying scalar field fluctuations $\pi$.     The violation of the consistency relations in this example can be traced to singularities in the Jacobian relating the path integral measure for $\pi$ and $\zeta$ fluctuations, as further explained in that section.

Conclusions, as well as a discussion of possible extensions of this work to tensors and to multi-field models are given in sec.~\ref{conclusions}.

\section{Formalism and Results}\label{formalism}

We work in the context of single-field models, in which case there is a gauge---which we refer to as  ``unitary gauge"---in which the physical scalar perturbations $\zeta(x)$ and traceless tensor modes $\gamma_{ij}(x)$ are encoded in the metric
\cite{Maldacena:2002vr}. Written in ADM form, this is
\begin{equation}
\label{eq:ADM}
ds^2 = -N^2 dt^2  + a^2(t)  h_{ij} (dx^i + N^i dt ) (dx^j + N^j dt) \; , 
\end{equation}
where
\be \label{h}
h_{ij}= e^{2\zeta} \left(e^\gamma\right)_{ij} =  e^{2\zeta} ( \delta_{ij} + \gamma_{ij}+\cdots) \; , \qquad \gamma_{ii}=0 \; .
\ee
The lapse $N$ and shift $N^i$ can be integrated out through the ADM Hamiltonian and momentum constraints (see~\cite{Maldacena:2002vr}). The physical tensor mode is then the transverse projection of $\gamma_{ij}$.    

 In this paper we focus on momentum-space scalar correlators at some fixed time $t_*$, $\langle\zeta(t_*)_{{\vec k}_1}\cdots \zeta(t_*)_{{\vec k}_n}\rangle$, with $\zeta_{\vec k}(t) = \int d^3 {x} \,  e^{-i{\vec k}\cdot {\vec x}} \zeta({\vec x},t)$.  For these, it is 
particular convenient to introduce a three-dimensional Euclidean generalization of the quantum effective action formalism of QFT (see e.g.~\cite{Weinberg2}).
We define a generating functional
\begin{equation} \label{Z}
Z[J] = \int D\zeta(t_*,{\vec x}) P[\zeta,t_*] \, e^{\int d^3 { x} \, \zeta({\vec x},t_*) J({\vec x})} \; ,
\end{equation}
where the probability measure $P[\zeta,t_*]$ on scalar modes at fixed time $t_*$ is an integral over the vacuum wave-functional $\Psi[\zeta,\gamma,t_*]$ at time $t_*$: 
\begin{equation}
\label{PPsi}
P[\zeta,t_*] = \int D\gamma_{ij}(t_*,{\vec x}) \, |\Psi[\zeta,\gamma,t_*]|^2 \; .
\end{equation} 
The wave-functional at time $t_*$ has the path integral representation
\begin{equation}
\label{eq:wotu}
\Psi[\zeta,\gamma,t_*] = \int^{\zeta(t_*),\gamma(t_*)} D\zeta(x) D\gamma_{ij} \, e^{iS[\zeta,\gamma]} \Psi_0[\zeta,\gamma] \; .
\end{equation}
Here, the initial wavefunction in the far past is $\Psi_0=\langle \zeta,\gamma;t\rightarrow -\infty|0\rangle$.   Although the initial state $|0\rangle$ is usually taken to be the Bunch-Davies adiabatic vacuum,  the specific form of the wavefunction will not matter for our results, as long as it is diffeomorphism invariant.   Note that we only integrate over $\zeta(x)$ and $\gamma_{ij}(x)$ as we assume that the non-dynamical fields $N,N^i$ have been integrated out via the ADM constraints.
Note also that the fact that $\Psi$ appears quadratically in Eq. (\ref{PPsi})
means the corresponding path integral has a doubling of fields -- this is precisely
the in-in formalism \cite{inin}.

Equal-time correlators can be retrieved from $Z[J]$ in the usual way, e.g.
\be
\langle\zeta(t_*)_{{\vec k}_1} \zeta(t_*)_{{\vec k}_2}\rangle =\frac{1}{Z[0]}
 \frac{\delta^2 Z[J]}{\delta J_{-\vec k_1} \delta  J_{-\vec k_2}}\bigg|_{J=0} \; .
\ee
In addition to the generating functional $Z[J]$, it is useful to introduce $W[J]=\ln Z[J]$, which generates connected correlators, as well as its Legendre transform 
\begin{eqnarray}
\label{Gamma}
\Gamma [\bar\zeta] = W[J] - \int d^3 {x} \, J(\vec x) \bar\zeta(\vec x) \; ,
\end{eqnarray}
which generates one-particle irreducible (1PI) correlators.   
\footnote{
More precisely, these are correlators with diagrams that are 1PI with respect to internal $\zeta$ lines, but not 1PI for the other fields.}
It follows from the definitions that
\be \label{zeta and J}
{\bar\zeta}(\vec x) = \frac{\delta W[J]}{\delta J(\vec x)} \; , \qquad J (\vec x)=- \frac{\delta\Gamma[{\bar\zeta}]}{\delta{\bar\zeta} (\vec x)}\; .   
\ee
In particular, $\bar \zeta(\vx)$ is the one-point function of $\zeta(\vx)$  for given external source $J(\vx)$:
\be \label{zeta bar}
\bar \zeta (\vx )= \langle \zeta (\vx)\rangle_J \; .
\ee 

The equal-time 1PI correlation functions are\footnote{We employ the momentum-space notations $\delta (\vk) \equiv (2\pi)^3 \delta^3(\vec k)$ and $\int_\vk \equiv \int 
\frac{d^3 k}{(2 \pi)^3}$.}
\be
\left. {\delta\over\delta \bar \zeta_{-\vec k_1}} \cdots  {\delta\over \delta \bar \zeta_{-\vec k_n}}  \Gamma[{\bar\zeta}]\right|_{{\bar\zeta}=0} =  \delta \Big(\sum_a {\vec k}_a \Big)   \Gamma^{(n)}({\vec k}_1,\cdots {\vec k}_n,t_*)  \; .
\ee 
The connected Green's functions generated by $W[J]$ can then be expressed in terms of the 1PI correlators $\Gamma^{(n)}({\vec k}_1,\cdots,{\vec k}_n,t_*)$.   For instance, the connected two-point function (or power spectrum), $\langle \zeta(t_*)_{\vk_1} \zeta(t_*) _{\vk_2}\rangle_c =\delta(\vk_1+\vk_2) \, G_c^{(2)}(\vk_1,\vk_2,t_*)$,  obtained by differentiation of $W[J]$, is
\begin{equation}
G_c^{(2)}(\vk,-\vk,t_*)=P(\vk,t_*)= -{1\over \Gamma^{(2)}(\vk,-\vk,t_*)} \; ,
\end{equation}
while the three-point function can be expressed as
\begin{equation}
\label{eq:3pt}
G^{(3)}_c({\vec k}_1,{\vec k}_2, {\vec k}_3,t_*)= \left[\prod_{a=1}^3 P({\vec k}_a, t_*)\right] \Gamma^{(3)}({\vec k}_1,{\vec k}_2, {\vec k}_3,t_*) \; .
\end{equation}

In general, the $\Gamma^{(n)}$ should be interpreted as quantum-corrected vertices which, combined with the equal-time propagators $P({\vec k}, t_*)$, yield the fully quantum-mechanical equal-time connected correlators via standard {\em tree-level} diagrammatics.  We will refer to $\Gamma[\bar \zeta]$ as the ``3D effective action", and to its Taylor coefficients $\Gamma^{(n)}(\vk_1, \dots, \vk_n)$ as the ``3D effective vertices", to remind ourselves that they contain information only about equal-time correlators.  From now on, we drop the explicit dependence on time as it is understood that all correlators are at time $t=t_*$.

Our strategy  then is quite simple: we will show that the 3D effective action is invariant under the residual diffs of \eqref{eq:ADM}, \eqref{h}. These act non-linearly on $\bar \zeta$. Upon expanding $\Gamma$ in powers of $\bar \zeta$, these non-linear symmetries yield relations between successive 3D effective vertices, $\Gamma^{(n)}$ and $\Gamma^{(n+1)}$, for all $n$'s---and therefore, between $n$- and $(n+1)$-point equal-time correlation functions. Since the parameterization \eqref{eq:ADM}, \eqref{h} admits residual diffs only at zero-momentum, these relations between $\Gamma^{(n)}$ and $\Gamma^{(n+1)}$ arise only when in the latter one of the momenta is ``soft", $\vq \to 0$.

To show the invariance of $\Gamma[{\bar\zeta}]$ under residual diffs, we follow closely the standard manipulations that yield the symmetries of the {\em standard} 4D quantum effective action, see e.g.~\cite{Weinberg2}. We assume that the integration measure---including the wave function $\Psi_0$---in Eq.~(\ref{eq:wotu}) is invariant under spatial diffeomorphisms at the fixed time $t_*$. This is actually a subtle assumption for non-linearly realized symmetries, like our diffs, or the global symmetries acting on Goldstone bosons. We briefly touch upon this subtlety in the Appendix. %For the moment, suffice it to say that it is harmless. 
Then, since the action $S[\zeta, \gamma]$ is  diff-invariant, the measure $D\zeta(t_*) P[\zeta,t_*]$ in \eqref{Z} is also invariant. For an infinitesimal diffeomorphism $\zeta\rightarrow \zeta +\Delta\zeta$ we have, after changing variables in the definition of $Z[J]$ and expanding to linear order in $\Delta\zeta$,
\begin{equation}
0=\int d^3{ x} \, J({\vec x}) \int  D\zeta(t_*,{\vec x}) P[\zeta,t_*]  \, e^{\int d^3{x} J \zeta} \, \Delta\zeta(t_*,\vec x) \; .
\end{equation}
Upon dividing by $Z[J]$, this result then implies
\begin{equation}
\label{eq:master}
\int d^3{x} \, \langle \Delta\zeta(t_*,{\vec x})\rangle_J  \, {\delta\Gamma[{\bar\zeta}]\over \delta {\bar\zeta}(t_*,\vec x)}=0 \; ,
\end{equation}
where we have used Eq.~\eqref{zeta and J}.  The above equation is a symmetry statement: it states  that $\Gamma [\bar \zeta]$ is invariant under the (infinitesimal) transformation
\be \label{sym}
\bar \zeta  \to \bar \zeta + \langle \Delta\zeta(t_*,{\vec x})\rangle_J \; ,
\ee
where $J$ and $\bar \zeta$ are related via \eqref{zeta and J}.
%We now consider the constraints on the form of correlators $\Gamma^{(n)}$ that follow from Eq.~(\ref{eq:master}).
In sec.~\ref{sec:dilations}  we consider the constraints on the form of correlators $\Gamma^{(n)}$ that follow from Eq.~(\ref{eq:master}) as a result of invariance under residual spatial dilations.   In sec.~\ref{sec:special}, we consider constraints from spatial conformal transformations.   This will require a discussion of possible modifications to Eq.~(\ref{eq:master}) due to shifts in tensor modes. 

\subsection{Dilations}\label{sec:dilations}

It is simplest to first consider scale transformations of the spatial coordinates.     Under the infinitesimal dilation ${\vec x}\rightarrow (1-\lambda){\vec x}$, the transformation of the spatial metric $g_{ij}=e^{2\zeta} h_{ij}$ on constant time hypersurfaces implies that
\begin{equation} \label{scale}
\Delta \zeta({\vec x}) = \lambda(1+{\vec x}\cdot\nabla\zeta({\vec x})) \; ,
\end{equation}
or in momentum space
\be
\Delta\zeta_{\vec k}=\lambda \, \delta({\vec k}) -\lambda\nabla_\vk\cdot(\vk \, \zeta_\vk) \; .
\ee
Since this transformation is at most linear in $\zeta$, it follows that $\langle\Delta\zeta\rangle_J=\Delta{\bar\zeta}$ (recall Eq.~\eqref{zeta bar}), and thus from Eq~(\ref{eq:master}),
\begin{equation}
{\delta\Gamma[{\bar\zeta}]\over \delta {\bar\zeta}_{\vq=0}}=-\int_{\vk} {\bar\zeta}_{\vk} (\vk\cdot \nabla_\vk) {\delta\Gamma\over\delta\zeta_{-\vk}}.
\end{equation}
This holds for a generic configuration $\bar \zeta_\vk$. We can Taylor-expand about $\bar \zeta = 0$.
Applying $n$ partial derivatives $\delta^n/\delta{\bar\zeta}_{-\vk_1}\cdots\delta{\bar\zeta}_{-\vk_n}$ to both sides, and evaluating them at $\bar \zeta = 0$, we obtain the relations
\begin{align}
 \frac{\delta^{n+1} \Gamma}{\delta \bar\zeta_{\vq=0} \, \delta\bar\zeta_{\vk_1} \cdots  \delta\bar\zeta_{\vk_n}} &  \bigg|_{\bar\zeta = 0} = \\
 & - \left(\sum_{a=1}^n \vec k_a \cdot \nabla_{\vec k_a}\right) 
     \frac{\delta^{n} \Gamma}{\delta\bar\zeta_{\vec
    k_1} \cdots \delta\bar\zeta_{\vec k_n}} \bigg|_{\bar\zeta=0} \; . \nonumber
\end{align}
Defining a dilation derivative ${\cal D}_n=\sum_{a=1}^n \vk_a\cdot\nabla_{\vk_a}$, it follows that  ${\cal D}_n \delta(\sum^n_{a=1} \vk_a) = -3$ and thus we can write the above equation as
\begin{equation}
\label{Gammamaster}
\Gamma^{(n+1)} (0, \vec k_1, \cdots, \vec k_n) = \left( 3 - {\cal D}_n
%  \sum_{i=1}^n \vec k_i \cdot \nabla_{\vec k_i} 
\right) \Gamma^{(n)} (\vec
  k_1, \cdots, \vec k_n) \, .
\end{equation}
Note that although the momenta on the RHS of this equation are constrained to add up to zero, the derivative operator ${\cal D}_n$ acts to generate an overall re-scaling that preserves the surface $\sum_a \vk_a=0$.  Consequently ${\cal D}_n$ is well defined when acting on $\Gamma^{(n)}$.
One can also think of 
$\Gamma^{(n)}$ as dependent on only $n-1$ momenta,
e.g. $\vec k_1, ..., \vec k_{n-1}$, and thus 
${\cal D}_n \Gamma^{(n)} = {\cal D}_{n-1} \Gamma^{(n)}$. 

We find that Eq. (\ref{Gammamaster}) is the most transparent way to state the dilation consistency relation.  Starting from this equation it is straightforward to derive constraints on the connected Green's functions
\begin{eqnarray}
\langle \zeta_{\vec k_1} ... \zeta_{\vec k_n} \rangle_c
=  \delta\Big(\sum_a\vk_a\Big) G_c^{(n)} (\vec k_1, \cdots,
\vec k_n).
\end{eqnarray}
For example, from  Eq. (\ref{Gammamaster}) with $n=2$ and Eq.~(\ref{eq:3pt}), we deduce that
\begin{eqnarray}
&& G_c^{(3)} (0, \vec k_1, \vec k_2) = - P(0) \left(3 + {\cal D}_1 \right) P(k_1) \nonumber \\
&& \quad = - P(0) P(k_1) {\partial {\,\rm ln} \big( k_1^3 P(k_1) \big) \over \partial{\,\rm
    ln} k_1} \, ,
\end{eqnarray}
reproducing Maldacena's consistency relation for the bispectrum, obtained here without making assumptions regarding the behavior of $\zeta$ at and after horizon crossing, or even the existence of an Hubble horizon.    

As another example, we may consider constraints on the four-point connected correlator.   Schematically, this can be expressed in terms 1PI correlators as
\begin{equation}
G_c^{(4)} = \bigg[\prod_a P(\vk_a)\bigg]  \bigg[ \Gamma^{(4)} +\sum_{s,t,u} \Gamma^{(3)} \, P(\vk) \,\Gamma^{(3)} \bigg],
\end{equation}
where the sum on the RHS is over all three ``crossings" of the external momenta.  Taking the limit $\vk_4\rightarrow 0$ on the LHS, it follows immediately from Eq.~(\ref{Gammamaster}) that
\begin{eqnarray}
G^{(4)}_c (0, \vec k_1, \vec k_2, \vec k_3) = - P(0)
(6 + {\cal D}_2) G^{(3)}_c (\vec k_1, \vec k_2, \vec k_3) \, ,
\end{eqnarray}
in agreement with the results of  ref.~\cite{Creminelli:2012ed}.     It is not difficult to show, by induction on $n$ that Eq.~(\ref{Gammamaster}) is in fact consistent with the results of~\cite{Creminelli:2012ed} for all possible connected Green's functions $G^{(n)}_c$:
$G^{(n+1)}_c (0, \vec k_1, ...)= -P(0) [3(n-1) + {\cal D}_{n-1}] G^{(n)}_c (\vec k_1, ...)$.

\subsection{Special conformal transformations}\label{sec:special}

We can repeat essentially the same analysis for infinitesimal special conformal transformations, $\delta \vx = \vx^2{\vec b} - 2 (\vec b\cdot \vx) \vec x$, under which
$\zeta$ shifts by
\begin{equation} \label{special}
\Delta\zeta({\vx}) = 2 \, {\vec b}\cdot \vx -\Big(\vx \, ^2 \,  {\vec b} - 2 (\vx\cdot{\vec b})\, \vx \Big)\cdot \vec \nabla \zeta(\vx) \; ,
\end{equation}
or in momentum space 
\begin{eqnarray}
\nonumber
  \Delta\zeta_{\vk}  &=& 2 i {\vec b}\cdot  \vec \nabla_{\vk} \delta(\vk) - i\nabla^2_\vk ({\vec b}\cdot \vk \, \zeta_\vk)\\
&& {} +2 i{\vec b}\cdot\nabla_\vk \left(\nabla_\vk\cdot (\vk \, \zeta_\vk)\right) \; . \label{delta zeta SC}
\end{eqnarray}

It should be noted however that, in the presence of the tensor mode $\gamma_{ij}$, this is {\em not} a residual gauge freedom of \eqref{eq:ADM}, \eqref{h}. More precisely, under a special conformal diffeomorphism, on top of the $\zeta$ shift just alluded to, one gets a transformation of $\gamma_{ij}$ that does not preserve its transversality. It should be possible to modify the following analysis to explicitly keep track of this. In this paper, we just ignore this subtlety and leave addressing it for future work. 
It suffices to note that the 1PI correlation functions involving $\zeta$ only
should not be affected by this subtlety. 
The reason is the following: Under a special conformal transformation, one gets a non-transverse contribution to $\gamma$ of order $\delta \gamma = {\cal O}(b \, \gamma)$. This contains a traceless scalar piece of the form $(\di_i \di_j - \sfrac13 \delta_{ij} \, \nabla^2) \chi$, which can be removed via a  further spatial diff, thus modifying eq.~\eqref{special} by an extra contribution of order $\chi = {\cal O}(b \, \gamma)$. It is important to notice that this extra diff is not simply undoing \eqref{special}. Then, if we generalized our formalism to include tensors in $\Gamma$, we would get a 
new symmetry statement replacing \eqref{sym}:
\be
\int d^3 x \, \frac{\delta \Gamma}{\delta \bar \zeta} \Delta \bar \zeta +  \frac{\delta \Gamma}{\delta \bar \gamma_{ij}} \Delta \bar \gamma_{ij} = 0
\ee
By taking functional derivatives with respect to $\bar \zeta$, and setting $\bar \zeta = \bar \gamma_{ij} =0 $---which is all we need for 1PI vertices of $\bar \zeta$ only---we see that these ${\cal O}(b \, \gamma)$ corrections to the tranformation laws simply do not contribute.
Since the 1PI $\zeta$-vertices are not affected, consistency relations at the level
of connected $\zeta$ Green's functions should thus remain valid as long as
diagrams where 1PI vertices are connected by internal graviton legs are subdominant.
%What this means is that the results that follow are valid only under the approximation that the tensor modes can be neglected, e.g., at lowest order in the tensor-to-scalar ratio for cosmological correlation functions of $\zeta$.

Inserting Eq.~\eqref{delta zeta SC} into Eq.~(\ref{eq:master})---for the same reason as before we have $\langle\Delta\zeta\rangle_J=\Delta{\bar\zeta}$---and differentiating $n$ times with respect to $\zeta$ then yields
\begin{align}
2 \nabla_{\vq\rightarrow 0}  \Big[\delta(\vq + & \sum_a \vk_a  ) \, \Gamma^{(n+1)}(\vq,\vk_1,\cdots,\vk_n)\Big] = \nonumber \\
& \sum_a \vec {\cal S}_a \Big[ \delta(\sum_a \vk_a)  \, \Gamma^{(n)}(\vk_1,\cdots,\vk_n) \Big] \; ,
\label{eq:mess}
\end{align}
where we have introduced the (vector) differential operator
\be
\vec {\cal S}_a \equiv \vk_a \nabla^2_{\vk_a} - 2 (\vk_a \cdot \vec \nabla_{\vk_a}) \vec  \nabla_{\vk_a} \; .
\ee

In order to convert Eq.~(\ref{eq:mess}) into a relation involving only gradients acting on $\Gamma^{(n+1)}$ and $\Gamma^{(n)}$, a prescription for evaluating derivatives $\nabla_{\vk_a}$ acting on functions on the surface $\sum_a \vk_a=0$ must be specified.  In other words: by construction the $G$'s and $\Gamma$'s are defined only on the surface $\sum_a \vk_a=0$; one can in principle continue them to arbitrary, unconstrained momenta, but there is no unique prescription for doing so. One could impose total symmetry under generic permutations of the momenta, but that is not enough to give an unambiguous continuation. For instance, away from $\sum_a \vk_a=0$, one can continue the inflationary spectrum $G^{(2)} \sim 1/k^3$ into a symmetric function of two independent momenta, but there are many inequivalent choices that reduce to the desired spectrum when these momenta are taken to be equal in magnitude and opposite in direction, e.g.~$1/(k_1+k_2)^3$ and $1/(k_1 k_2)^{3/2} $. Generic derivatives with respect to the momenta take us (infinitesimally) away from the $\sum_a \vk_a=0$ surface, and are thus sensitive to which choice we adopt.
%\footnote{This subtlety was irrelevant for the case of dilations, because the combination ${\cal D} = \sum \vk_a \cdot \vec \nabla_{\vk_a}$ preserves the  $\sum_a \vk_a=0$ constraint.}
As we show in Appendix \ref{ambiguity}, it is straightforward to check that in the end these ambiguities cancel between the left hand side and the right hand side of the final consistency relation, provided one preserves the {\em dilation} consistency relations in moving off the momentum-conserving surface. However, instead of having a consistency  relation with equally ambiguous sides, we prefer to avoid the ambiguity altogether. As we will see in the next section, our pickiness about this will be rewarded.

One possible choice is to eliminate, e.g., $\vk_n$ using the delta functions, so that $\Gamma^{(n)}$ is only a function of the $n-1$ momenta $\vk_{a=1,\cdots,n-1}$, and $\Gamma^{(n+1)}$ of those as well as of $\vk$.  This has the advantage of expressing the consistency conditions directly in terms of observationally relevant quantities, such as the spectrum, the bispectrum, etc.  With this prescription in mind, the LHS of Eq.~\eqref{eq:mess} can then be expressed as
\be
\label{eq:lhs} 2 \, \delta(\sum_a \vk_a) \nabla_{\vq\rightarrow 0} \Gamma^{(n+1)} 
+ 2  \Big[\nabla_{\vq\rightarrow 0}\delta(\vq+\sum_a \vk_a) \Big](3-{\cal D}_n) \Gamma^{(n)} 
\ee
where we have made use of the dilation consistency relation, Eq.~(\ref{Gammamaster}).

The RHS of Eq.~(\ref{eq:mess}) is a sum of terms with up to two derivatives acting on $\delta(\sum_a\vk_a)$.   The term with two gradients on the delta function is 
\be
\label{eq:d2d}
\Gamma^{(n)} \, \sum_a \vec {\cal S}_a \, \delta(\sum_b \vk_b) = 
 6  \Big[\nabla_{\vq\rightarrow 0}\delta(\vq+\sum_a \vk_a)\Big] \Gamma^{(n)}
\ee
where we have made use of the identity
\footnote{Such an identity can be proved via standard manipulations upon rewriting the delta function as $\delta(\sum_b \vk_b) = \int d^3 x \, e^{-i (\sum_b \vk_b)\cdot \vec x}$.}
\begin{align}
\sum_a \vk^i_a & \nabla^j_{\vk_a} \nabla^k_{\vk_a}  \delta(\sum_b \vk_b) =\\
&   -(\delta^{ij} \nabla^k_{\vq\rightarrow 0}+\delta^{ik} \nabla^j_{\vq\rightarrow 0}) \, \delta(\vq+\sum_a \vk_a) \; . \nonumber
\end{align}
The term involving a single gradient acting on the delta function is
\begin{eqnarray}
\label{eq:crossterm}
%\nonumber
%2  \sum_i \left[\left(\vk^a _i \nabla_{\vk_i}^b - \vk^v _i \nabla_{\vk_i}^a\right)\Gamma^{(n)}\right] \nabla^b_{\vk_i} \delta(\sum_i \vk_i)\\
%\nonumber-2 \sum_i (\vk_i\cdot\nabla_{\vk_i})\Gamma^{(n)} \nabla^a_{\vk_i} \delta(\sum_j\vk_j)=\\
2\Big[\big( L^{ij} - \delta^{ij} {\cal D}_n \big) \Gamma^{(n)}\Big] \Big[\nabla^j_{\vq\rightarrow 0} \delta(\vq+\sum_a \vk_a) \Big] \; ,
\end{eqnarray} 
where $L^{ij} \equiv \sum_a (\vk^i _a \nabla_{\vk_a}^j  - \vk^j _a \nabla_{\vk_a}^i)$ is the angular momentum generator. By
rotational invariance, we have $L^{ij} \Gamma^{(n)}=0$, so Eq.~(\ref{eq:crossterm}) reduces to 
\be
-2  \Big[\nabla_{\vq\rightarrow 0} \delta(\vq+\sum_a \vk_a) \Big] {\cal D}_n \Gamma^{(n)} \; .
\ee   
Finally the term with no gradients acting on the delta functions is simply
\begin{eqnarray}
\label{eq:d0d}
\delta(\sum_a \vk_a) \sum^{n-1}_{a=1} \vec {\cal S}_a \,  \Gamma^{(n)}
\end{eqnarray}
where we have interpreted $\Gamma^{(n)} = \Gamma^{(n)}(\vk_1,\cdots,\vk_{n-1},-\sum_{a=1}^{n-1}\vk_a)$ as discussed previously.

Comparing the various terms in Eqs.~(\ref{eq:d2d}), (\ref{eq:crossterm}), (\ref{eq:d0d}) with Eq.~(\ref{eq:lhs}), we find that all terms involving the gradient  $\nabla_{\vq\rightarrow 0}\delta(\vq+\sum_i \vk_i)$, cancel, so that Eq.~(\ref{eq:mess}) finally becomes 
\begin{align}
\label{eq:scresult}
2 \nabla_{\vq\rightarrow 0}   \Gamma^{(n+1)} & (\vq,\vk_1,\cdots,-\vq-\sum_{a=1}^{n-1}\vk_a) = \\
\nonumber
&\sum_{a=1}^{n-1} \vec {\cal S}_a \,  \Gamma^{(n)}(\vk_1,\cdots,\vk_{n-1},-\sum_{a=1}^{n-1}\vk_a).
\end{align}
This result can be combined with its
counterpart for dilation, Eq. (\ref{Gammamaster}), into
a Taylor expansion in $\vec q$:
\begin{equation}
\label{TaylorGamma}
\Gamma^{(n+1)} (\vq, \vec k_1, ...)
= \left(3 - {\cal D}_n + {1\over 2} \sum_{a=1}^{n-1} \vec q \cdot \vec S_a
\right) 
\Gamma^{(n)}(\vk_1, ...) \, ,
\end{equation}
with corrections coming at $O(q^2)$.

%relevant notes: JW p. 113 - 117, JustinC p. 187 - 189
It is straightforward to convert Eq. (\ref{eq:scresult}) into a statement regarding the connected Green's functions $G_c^{(n)}$.   For example, for $n=2$ we obtain
\begin{align}
\nonumber
\label{n=2SCF}
\nabla_{\vq\rightarrow 0} G^{(3)}(\vq,\vk_1,-\vec q -\vk_1) & =
- {\nabla_{\vq\rightarrow 0}P(\vq)} (3 + {\cal D}_1) P(\vk_1) \\
& - \sfrac12  P(0) \left( 6 \nabla_{\vec k_1} - \vec {\cal S}_1 \right) \,  P(\vk_1)
\, .
%& = \dots
\end{align}
In deriving this, it is important to keep in mind $G^{(3)}(\vec q, \vec k_1,
- \vec q - \vec k_1) = \Gamma^{(3)} (\vec q, \vec k_1,
- \vec q - \vec k_1) P(\vec q) P(\vec k_1) P(\vec q + \vec k_1)$,
and thus its derivative with respect to $\vec q$ has
several terms, including the derivative of $P(\vec q + \vec k_1)$
which is easy to miss.
Eq. (\ref{n=2SCF})---at first sight---looks the same as the consistency relation derived in \cite{Creminelli:2012ed}. There is one important difference however: our $\vk_n$ momentum has been expressed in terms of the other momenta, as enforced by the delta-functions, {\em before} taking the derivatives.   As the example in sec.~\ref{sec:exsr} shows, this apparently minor technicality has crucial consequences.

\section{Examples}
\label{sec:examples}

\subsection{Slow-roll inflation}
\label{sec:exsr}

It is well known that the three-point correlator in single-field slow-roll inflation obeys the relation in Eq.~(\ref{eq:1}),  which in our language is associated with residual dilations. What has been overlooked so far is that it also obeys a  non-trivial relation associated with special conformal transformations.   More precisely:  the special conformal relation between two-point and three-point correlators has been claimed to be trivially obeyed~\cite{Creminelli:2012ed}, with the left- and right-hand sides both being zero in the squeezed limit.  In fact, we claim that this is an artifact of insisting on using (somewhat ambiguous) $n$-point functions that depend on $n$ unconstrained momenta.   If this ambiguity is resolved by a prescription such as the one discussed above, and one expresses the $n$-point correlators in terms of $n-1$ independent momenta, there is a non-trivial check to perform already at the level of the three-point function. 

The three-point function has been computed by Maldacena \cite{Maldacena:2002vr}. At lowest order in the slow roll expansion, it reads
\begin{align}
G^{(3)}& (\vk_1,\vk_2,\vk_3)  \simeq \frac{H^4}{4 \epsilon^2 M_{\rm Pl}^4 } \frac{1}{\Pi_i (2 k_i^3)} \times \\
& \Big[ (3 \epsilon - 2\eta) \sum_i k_i^3 + \epsilon \sum_{i \neq j} k_i k_j^2 + 8 \epsilon \frac{1}{\sum_i k_i} \sum_{i<j} k_i^2 k_j^2 \Big]
\; , \nonumber
\end{align}
while the spectrum is
\be \label{spectrum}
P(k) = -\frac{1}{\Gamma^{(2)}(k)} \simeq \frac{H^2}{4 \epsilon \, M_{\rm Pl}^2} \frac{1}{k^{3-2(\eta-3 \epsilon)}}
\ee
By setting $\vk_2 = -(\vk_1+\vk_3)$ and taking the soft limit $\vk_3 \to 0$, after straightforward manipulations we get
\begin{align}
\Gamma^{(3)} & = \prod_a\big (-\Gamma^{(2)}(k_a) \big) \cdot G^{(3)} \\
& = (\eta- 3 \epsilon) \Gamma^{(2)}(k_1) \left[ 2+ 3  \frac{\vk_1 \cdot \vk_3}{k_1^2}\right] + {\cal O}(k_3^2) \; .
\nonumber
\end{align}
The zeroth order term obeys the dilation consistency condition:
\be
\Gamma^{(3)}\big|_{\vk_3 \to 0} = 2 (\eta-3\epsilon) \Gamma^{(2)}(k_1) = (3-{\cal D}_1)\Gamma^{(2)}(k_1) \; ,
\ee
in agreement with the standard results.  What is new here is that the linear term in $\vk_3\rightarrow 0$ obeys the special-conformal consistency condition:
\begin{align}
2 \nabla_{\vk _3 \rightarrow 0}  \Gamma^{(3)}  & = 6(\eta-3 \epsilon) \frac{\vk_1}{k_1^2} \cdot \Gamma^{(2)}(k_1) \\
& = \vec {\cal S}_{\vk_1}  \, \Gamma^{(2)}(k_1) \nonumber \; ,
\end{align}
as can be checked straightforwardly from the explicit form of the spectrum, Eq.~\eqref{spectrum}.

\subsection{Flat-space limit of ghost condensate}
\label{ghost}

A less traditional example is the ghost condensate in Minkowski spacetime, coupled to dynamical gravity.
This is the theory of a peculiar derivatively coupled scalar field $\phi$. At lowest order in the derivative expansion, in addition to the gravitational (Einstein-Hilbert) term, the theory is defined by the Lagrangian~\cite{ArkaniHamed:2003uy}
\begin{equation}
{\cal L} = \sqrt{-g} M^4 P(X) \; , 
\end{equation}
with $X=-g^{\mu\nu} \partial_\mu \phi \partial_\nu \phi$, $M$ some mass scale, and $P$ a fairly generic function. If such a function has a minimum at some $X$---say $X=1$---and if the value of $P$ at that minimum is adjusted to be zero (by tuning the cosmological constant), then
\be \label{gc}
\phi(x) = t \; , \qquad g_{\mu\nu} (x) = \eta_{\mu\nu}
\ee 
is a solution of the equations of motion {\em in the presence of gravity}.      
%If we tune the zero of energy by the condition $P(1)=0$, the metric corresponding to the solution $\phi(x) = t$ is Minkowski space.
This model thus shares with single-field inflationary scenarios the presence of a ``rolling" physical scalar that can serve as a clock---thus allowing us to choose the unitary gauge that we have introduced in sect.~\ref{formalism}---, but does not feature a cosmological expansion nor an Hubble horizon. In particular, modes do not ``freeze-out" at late times, nor do they become classical.

Before proceeding, it should be mentioned that higher-derivative terms in the action are crucial in order to stabilize the solution $\phi = t$ against small perturbations:   from the $P(X)$ Lagrangian by itself, perturbations do not have gradient energy, which results---neglecting gravity---in a trivial dispersion relation, $\omega = 0$. When higher derivative terms are included, the leading gradient energy gives the perturbations a quadratic dispersion relation at short distances, 
\be \label{omega_ghost}
\omega_{\vec k} \simeq  \frac{\bar M}{M^2}k^2 \; ,
\ee
where $\bar M$ is a mass parameter associated with the higher-derivative terms. In the far infrared, at momenta lower than
\be
k_J = \frac{M^2}{M_{\rm Pl}} \; ,
\ee
gravitational effects become important, and a slow, Jeans-type instability sets in \cite{ArkaniHamed:2003uy}. To be safe, we will work at shorter distances, $k \gg k_J$, where the solution \eqref{gc} is free of instabilities
\footnote{We  work in the weak gravity limit, $\mpl \gg \bar M, M$. For simplicity we can also assume $\bar M \sim M$. These conditions automatically ensure  that the Jeans scale $k_J$ is much below the strong-coupling scale of the theory, which is some combination of $M$ and $\bar M$.}.

The perturbations about cosmological solutions of this theory were analyzed in ref.~\cite{Creminelli:2006xe}, working in a unitary gauge where all fluctuations appear in the metric, written in the ADM parametrization employed in Eq.~(\ref{eq:ADM}).    The action for the scalar perturbation $\zeta$ is quite involved, see~\cite{Creminelli:2006xe} for details.    We will use those results (and extend them to cubic order) for the case where the background metric is Minkowski.    At high momenta, $k \gg k_J$, the quadratic Lagrangian reads
\begin{equation}
\label{eq:l2}
{\cal L}_2 \simeq {2 M^4_{\rm Pl}\over {\bar M}^2} {\dot \zeta}^2  - {2 M^4_{\rm Pl}\over M^4} (\nabla^2\zeta)^2 ,
\end{equation}
while the cubic interactions relevant for checking the three-point  consistency relations (i.e., those that survive when one of the three momenta becomes soft) are given by
\footnote{In the notation of ref.~\cite{Creminelli:2006xe}, we have used the following unitary-gauge Lagrangian for the ghost condensate
\be
S = \int d^4x \sqrt{h} \big[ \sfrac18 M^4 \,(1/N^2-1)^2 - \sfrac12 \bar M^2 (E^i {}_i)^2 \big] \; ,
\ee
which is not the most general one. However, for the purposes of our check we need not be completely general---just consistent.}
\begin{eqnarray}
{\cal L}_3 \simeq  {6 M^4_{Pl}\over {\bar M}^2} \zeta \dot\zeta^2 +  {2M^4_{Pl}\over M^4} \zeta (\nabla^2\zeta)^2 \; .
\end{eqnarray}

The time-ordered two-point function of $\zeta$, from Eq.~(\ref{eq:l2}) is
\begin{eqnarray}
\nonumber
&& \langle T \zeta_{\vk}(t) \zeta_{\vec q}(t')\rangle = \delta(\vk+{\vec q}) \, {{\bar M}^2 \over 8 M^4_{Pl}  \omega_\vk}\\
&&\times  \left[\theta(t-t') e^{-i\omega_\vk (t-t')}+\theta(t'-t) e^{-i\omega_\vk (t'-t)}\right],
\end{eqnarray}
with $\omega_{\vk}$ given in \eqref{omega_ghost}, so that
\be
 \Gamma^{(2)}(\vk,-\vk) = -{8 M_{Pl}^4\over {\bar M}^2} \omega_{\vk} \; ,
\ee
while the cubic $\zeta$ self-interaction with incoming momenta $\vk_{1,2,3}$ can be written as 
\begin{eqnarray}
-{8iM^4_{Pl}\over {\bar M}^2}(\omega_{\vk_1} \omega_{\vk_2}+\omega_{\vk_1} \omega_{\vk_3}+\omega_{\vk_2} \omega_{\vk_3}),
\end{eqnarray} 
up to contact terms that do not contribute to long distance correlations (an overall delta function $\delta(\sum_a \vk_a)$ has been suppressed).    The connected three-point function at equal time $t$ is therefore
\begin{eqnarray}
\nonumber
 G^{(3)}_c(\vk_1,\vk_2,\vk_3,t) &=& -{8iM^4_{Pl} \over {\bar M}^2}  \left( {\bar M}^2 \over 8 M^4_{Pl}\right)^3  \left[\sum_a \omega_{\vk_a}^{-1}\right] \\
 &&{} \times {}{\cal I}(\omega_a,t),
\end{eqnarray}
where ${\cal I}(\omega_a,t)$ is the integral
\begin{eqnarray}
{\cal I}(\omega_a,t) &=& \int_{-\infty}^\infty dt' \left[\theta(t-t') e^{-i\sum_a\omega_{\vk_a} (t-t')}\right.\\
\nonumber
&&\left.{} +\theta(t'-t) e^{-i\sum_a\omega_{\vk_a} (t'-t)}\right]= {2 i\over -\sum_a \omega_{\vk_a} + i\epsilon}.
\end{eqnarray}
The connected Green's function is then
\begin{eqnarray}
G^{(3)}_c(\vk_1,\vk_2,\vk_3,t) = - {16 M^4_{Pl} \over {\bar M}^2}  \left( {\bar M}^2 \over 8 M^4_{Pl}\right)^3 {\sum_a \omega_{\vk_a}^{-1}\over \sum_a \omega_{\vk_a}},
\end{eqnarray}
while the 1PI correlator at time $t$ is, from Eq.~(\ref{eq:3pt}),
\begin{equation}
\Gamma^{(3)}(\vk_1,\vk_2,\vk_3) =  - {16 M^4_{Pl} \over {\bar M}^2}  {\omega_{\vk_1} \omega_{\vk_2}+\omega_{\vk_1} \omega_{\vk_3}+\omega_{\vk_2} \omega_{\vk_3}\over  \omega_{\vk_1} + \omega_{\vk_2} + \omega_{\vk_3}}
\end{equation}
In the limit $\vk_3\rightarrow 0$ this becomes 
\begin{equation}
\Gamma^{(3)}(\vk,-\vk,0)=  -{8 M_{Pl}^4\over {\bar M}^2} \omega_{\vk}= \Gamma^{(2)}(\vk,-\vk) .
\end{equation}
Given that $(3-{\cal D}_1) \omega_{\vk} = \omega_{\vk}$, this verifies the dilation consistency relation for the three-point function. 
If before taking the soft limit $\vk_3 \to 0$, we differentiate with respect to $\vk_3$, we can check the special conformal consistency condition as well:
\begin{align}
2 \nabla_{\vk _3 \rightarrow 0} & \Gamma^{(3)}\big( \vk,-(\vk+\vk_3), \vk_3 \big)  = - 16 \frac{M^4_{\rm Pl}}{\bar M M^2} \vk  \\
& =\vec {\cal S}_\vk  \, \Gamma^{(2)}(\vk, -\vk) \; ,  \nonumber
\end{align}
in agreement with Eq.~\eqref{eq:scresult}.

Because the background is flat, this example serves to illustrate that the consistency relations hold not as a consequence of super-horizon freeze-out of modes but rather because of the residual diffeomorphisms associated with scale and special conformal invariance.

\subsection{Non-example:   flat space scalar field}\label{counter}

Despite the relatively few assumptions that go into our derivation of the consistency relations, our results are not completely model independent.   A simple example that violates the consistency relations is the theory of a free massless scalar field $\phi(x)$ in flat spacetime---decoupled from gravity---with a time-dependent background.   In a coordinate system $(x^0,{\vec x})$ in which the metric is Minkowskian, the theory has a solution with $\phi(x) = M^2 x^0$ for some arbitrary scale $M$, and fluctuations about this background are parametrized as $\phi(x)=M^2(x^0+\pi(x))$.    

It is also possible to describe the physics in `unitary gauge'  with coordinates $(t,\vec X)$ in which $\phi$ is spatially homogeneous on surfaces of constant time $t$,
\begin{eqnarray}
t(x^0,{\vec x}) &=& x^0+\pi(x^0,{\vec x}),\\
{\vec X}(x^0,{\vec x}) &=& {\vec x}.
\end{eqnarray}
We will need the following metric components in these coordinates
  \begin{eqnarray}
g^{tt}(t,{\vec X}) &=& -  {1\over \left(1-{\dot\pi}\right)^{2}}\left[1-  (\nabla\pi)^2\right],\\
\label{eq:smetric}
g_{ij}(t,{\vec X}) &=& \delta_{ij} + \nabla_i \pi \nabla_j \pi,
 \end{eqnarray}
 where $\pi$ is now regarded a function of $(t,{\vec X})$ via $\pi(t,{\vec X})\equiv \pi(x^0(t,{\vec X}),{\vec x}(t,{\vec X}))$ and ${\dot {} }=\partial/\partial t$, $\nabla_i=\partial/\partial X^i$.  Thus in these coordinates the action for the fluctuation $\pi$ becomes
 \begin{eqnarray}
\label{eq:piaction}
\nonumber
S[\pi]&=& -{M^4\over 2} \int dt d^3 {\vec X} (1-{\dot \pi}) g^{tt}(t,{\vec X})\\
&=& {M^4\over 2} \int dt d^3 {\vec X} \left[{\dot \pi}^2 - (\nabla\pi)^2\right]+{\cal O}(\pi^3)
\end{eqnarray}

To extract $\zeta$ from Eq.~(\ref{eq:smetric}), we decompose $g_{ij} = e^{2\zeta} h_{ij}$, with $h_{ij}$ unimodular, and perform a further diffeomorphism ${\vec X}\rightarrow {\vec X} + {\vec\xi}(t,{\vec X})$ such that $\gamma_{ij} = h_{ij}-\delta_{ij}$ is transverse, $\nabla_i \gamma_{ij}=0$, in the new coordinates .   To quadratic order in $\pi$, this requires
\begin{equation}
\sfrac{4}{ 3} \nabla\cdot{\vec \xi} =  {1\over\nabla^2} \nabla_i \nabla_j (\nabla_i\pi \nabla_j\pi) -\sfrac{1}{3} (\nabla\pi)^2+{\cal O}(\pi^3),
\end{equation}
yielding
\begin{equation} \label{zeta of pi}
4 \zeta =(\nabla\pi)^2 - {1\over\nabla^2} \nabla_i \nabla_j (\nabla_i\pi \nabla_j\pi) + {\cal O}(\pi^3).
\end{equation}

Seeing that the consistency relations do not hold in this example is now a matter of simple power counting in the scale $M$.    From Eq.~(\ref{eq:piaction}), the connected two-point functions is $\langle\zeta\zeta\rangle\sim M^{-8}$, and thus $\Gamma^{(2)}({\vec k},-{\vec k})\sim M^8$.   On the other hand, the connected three-point function scales as $M^{-12}$ so that $\Gamma^{(3)}\sim M^{12}$ up to terms involving more powers of momenta divided $M$.    Barring accidental cancellations, it follows that no simple relation between $\Gamma^{(2)}$ and $\Gamma^{(3)}$ is possible in this model.

In this example, there is a subtlety in applying the formal path integral manipulations of sec.~\ref{formalism}.   Although we can choose the standard unitary gauge of Eq.~\eqref{h}, the mapping between the scalar perturbation $\pi$ that appears in the original Lagrangian and the $\zeta$ variable defined in unitary gauge starts at quadratic order in the $\pi$ field. This is quite an unusual change of field variables for a perturbative field theory---quite different from the standard field redefinitions we are used to in quantum field theory, which, even if non-linear, always start with linear terms. Usually we require such linear + higher order field redefinitions because they do not affect $S$-matrix elements. In our case we are not computing $S$-matrix elements, and such a requirement should not be relevant. Perhaps more to the point, our $\zeta = {\cal O}(\pi^2)$ change of variables is not invertible in any small neighborhood of $\pi = \zeta =0 $, and it has a singular $\delta \pi/\delta \zeta$ Jacobian about that point. Since our theory is perturbative in $\pi$ (free, in fact) about $\pi=0$, such a change of integration variables in the path-integral is clearly pathological. We believe this to be the cause of the violation of the consistency relations in this example.

\section{Discussion and outlook}
\label{conclusions}
We derived fully quantum-mechanical consistency relations---or, in the language of quantum field theory, soft-$\zeta$ theorems---for 1PI vertex functions of scalar perturbations in cosmology. Our results follow from the residual gauge invariance of an equal-time analog of the quantum effective action for $\zeta$, which generates equal-time correlation functions. The near-deSitter isometries of inflation play no role in our derivation, nor do the existence of a cosmological horizon, the freezing-out of long-wavelength perturbations, or their becoming approximately classical. As a result, our consistency relations hold as exact quantum mechanical statements, and in more general situations than the standard inflationary cosmology scenarios, like the flat-space ghost-condensate example of sect.~\ref{ghost} shows. 

The residual gauge invariance we make use of is the three-dimensional conformal group, under which $\zeta$ transforms as a dilaton. For dilations, the presence of tensor modes does not affect our arguments, and our consistency relations can be  generalized straightforwardly to include tensor modes on the external legs. On the other hand, special conformal transformations do not preserve the transversality of tensor modes, and must be dealt with more carefully. For 1PI correlation functions that only involve zeta on the external legs, this does not constitute a problem, for the reasons outlined in sect.~\ref{sec:special}. For 1PI correlation functions that also have external tensors, this subtlety has to be addressed directly, perhaps via transverse projectors acting on the external tensors\footnote{A different approach will appear in \cite{HHK}.}.  We leave investigating this subtlety for future work.

Dilations and special conformal transformations have also been used recently in \cite{Creminelli:2012ed} to derive, via more traditional methods, consistency relations which---not surprisingly---are in agreement with ours.  However, we feel that the novelty of our work lies in clarifying the wide scope of validity of such consistency relations, and the minimality of the assumptions that go into deriving them.  Moreover, we identified (and removed) an implicit ambiguity in the special-conformal consistency relations.  Our unambiguous version can be tested non-trivially already at the level of the slow-roll inflation three-point function.

Given how few assumptions we have made in deriving the consistency relations, it is not completely obvious to see what goes wrong in situations where they are known to fail, like for example multi-field inflation. There, we can still choose a gauge in which one of the time-dependent scalars has zero fluctuations.   In that case, one can then parameterize the scalar modes in terms of the $\zeta$ field appearing in the metric in this gauge, and of the remaining scalar fields. All these modes will transform in some definite way under our residual gauge transformations, and, once fields other than $\zeta$ have been integrated out, our $\Gamma[\bar \zeta]$ should have the same (gauge) symmetries as we have used, and should thus obey the same consistency relations as we have derived. 

The catch in this argument is that this definition of $\zeta$ does not coincide with the observationally relevant late-time curvature perturbation:  all multi-field examples that violate the consistency relations, do so thanks to a substantial conversion of ``isocurvature" fluctuations into adiabatic ones. One might try to work around this by  choosing a `late-time unitary gauge', where the late-time curvature perturbations are encoded  in the spatial metric as in  \eqref{h}, and extrapolating this gauge choice backwards in time. It is not clear how this could work in general though: the most natural and universal definition of a post-reheating unitary gauge is to define time via the $\rho = {\rm const}$ hypersurfaces, but for our purposes this is not quite the same as unitary gauge. This is evident if one employs a $P(X)$ parameterization of the post-inflationary (vorticity-free) cosmic fluid, where $X = (\di \phi)^2$ for some scalar $\phi$, and $P$ is a generic function (related to the fluid's equation of state.) In this parameterization, $\rho = {\rm const}$ is the same as $X= {\rm const}$, which is {\em not} the same as $\phi = {\rm const}$. In other words, in the $\delta \rho = 0$ gauge, $\delta \phi$ is not zero, and scalar perturbations are parameterized by $\zeta$ {\em and} $\delta \phi$. The real unitary gauge choice would be to choose $\delta \phi =0 $ and define scalar perturbations in terms of the $\zeta$ variable appearing in the metric in that gauge, but it is not obvious how to extrapolate the $\delta \phi = 0 $ gauge condition to earlier times, i.e.~to inflationary times, when this scalar field $\phi$ has no meaning in general. 

A subtle paradox remains though, at least for models in which all isocurvature perturbations  eventually disappear. One could choose the {\em initial} time in our effective action manipulations to be in fact quite late in the evolution the universe, say {\em after} the isocurvature perturbations are gone. Then, one would be left with adiabatic scalar perturbations in the post-inflationary cosmic fluid, which can be parameterized in the $\delta \phi =0$ unitary gauge above by $\zeta$ only. Now there is no need to extrapolate this gauge back in time. It seems to us that in this case $\Gamma[\zeta]$ should obey all our symmetries, and should thus respect the consistency relations. We are actually unclear about why it does not.

Similar considerations apply to the case of ``solid inflation", whose three-point function also violates the consistency relations  \cite{ENW}. There, there is no standard unitary gauge choice: once the fluctuations in the matter fields are set to zero via a gauge choice, the metric cannot be put into the form  \eqref{h}. In fact, the most convenient version of unitary gauge in that case would be to {\em remove} $\zeta$ from the metric via a time-redefinition, and parameterize scalar perturbations in terms of a scalar component of the traceless $\gamma_{ij}$, which in this gauge cannot be taken as transverse \cite{ENW}. Like in the multi-field case, here too there is no gauge in which the curvature perturbation that is relevant for observations at late times appears in the spatial metric as $\zeta$ does in  \eqref{h}.  Finally, there is the counterexample of sec.~\ref{counter}, which despite the fact that it admits a standard unitary gauge violates the consistency relations.   In this case, the violation can be traced to the specific form of the relation between $\zeta$ and the underlying scalar fluctuation, as discussed in sec.~\ref{counter}.

\vspace{0.2cm}
 
\noindent
{\em Acknowledgements.}
We thank Paolo Creminelli,  Jorge Nore\~na, Marko~Simonovi\'c, and especially 
Justin Khoury and Kurt Hinterbichler for extensive discussions and
collaboration. We are grateful to Sergei Dubovsky, Raphael Flauger,
Alan Guth, Enrico Pajer and Matias Zaldarriaga for useful discussions.
This work  is supported by the DOE under contracts DE-FG02-11ER41743, DE-FG02-92-ER40699,
DE-FG02-92-ER40704, and 
by NASA under contract NNX10AH14G.
 
\appendix

\section{Non-linear symmetries of $\Gamma$}

The standard manipulations that are normally used to prove the symmetries of the effective action $\Gamma$ are not entirely correct when applied to non-linearly realized symmetries, like for instance spontaneously broken global symmetries. The reason is that the action of such  symmetries on the field variables in the path-integral changes the boundary conditions for those fields, e.g.~from zero to something non-zero for  Goldstone fields. Or, in the language of sect.~\ref{formalism}, the vacuum wave-functional $\Psi_0$ is {\em not} invariant under non-linearly realized symmetries: for spontaneously broken global symmetries, this is true {\em by definition}.
Since our residual diffs \eqref{scale} and  \eqref{special} formally act non-linearly on $\zeta$ (even though they do not correspond to spontaneous breaking in any standard sense), we have to keep this subtlety into account.

For non-linearly realized symmetries, to see any invariance at the level of the vacuum wave-functional we have to keep into account that at any fixed time $t$, the wave functional depends not only on the field configuration $\zeta(\vec x)$ , but  secretly also on a background configuration $\zeta_b(\vec x)$, which can be thought of as the boundary condition for $\zeta(\vec x)$ at spatial infinity, or more in general as the vev $\langle \zeta (\vec x)\rangle$ in the absence of external sources (the necessity for an $\vec x$ argument for $\zeta_b$ will be clear in a moment). The standard choice for the vacuum would be $\zeta_b = 0$, but a non-linear symmetry transformation can change that, so that the general symmetry statement for the vacuum wave-functional is
\footnote{For spontaneously broken symmetries, this property of the wave functional is just the statement that there are many physically equivalent vacua, all related by symmetry transformations, and the wave functionals of the fields in all these vacua are the same, provided one transforms the fields properly.}
\be \label{WF symmetry}
\Psi_0[\zeta , \zeta_b = 0 ] = \Psi_0[\zeta + \Delta \zeta, \zeta_b = \langle \Delta \zeta \rangle] \; .
\ee
For notational simplicity, from now on we will be ignoring the tensor modes, and we will omit the ``observation time'' $t_*$.  
% We will also ignore the subtlety that for generic non-linear transformations 
The symmetry transformations we will be interested in are the scale transformation \eqref{scale}, which generates a constant background
\be
\zeta_b = \lambda \; ,
\ee
and the special conformal transformations \eqref{special}, which generate a linear one,
\be
\zeta_b (\vec x) = \vec b \cdot \vec x \; . \label{x dep bkgrd}
\ee
One can now go through the same steps as in sect.~2, keeping the background-dependence of the various quantities explicit:
\begin{align}
\Psi[\zeta]  & \to \Psi[\zeta, \zeta_b] \; , \qquad
P[\zeta] \to P[\zeta, \zeta_b] \; , \\
 W[J]  & \to W[J, \zeta_b] \; , \qquad
\Gamma[\bar \zeta] \to \Gamma[\bar \zeta, \zeta_b] \; ,
\end{align}
where $\Gamma$ is still defined as the Legendre transform of $W$ with respect to~$J$ only.  In particular, standard properties of the
Legendre transform imply
\be
\frac{\delta \Gamma}{\delta \zeta_b} \Big|_{\bar \zeta} = \frac{\delta W}{\delta \zeta_b} \Big| _{J} \; .
\ee
%In this equation we use partial derivatives rather than functional ones, to emphasize that even though $\zeta_b$ can have some spatial dependence like in \eqref{x dep bkgrd}, it still depends on a finite number of parameters---$\lambda$ and $\vec b$ in our case---since it is generated by the action of a finite-dimensional symmetry group on the $\zeta_b=0$  configuration. So, the partial derivatives above should really be thought of as derivatives with respect to these parameters.

Using the symmetry statement \eqref{WF symmetry},  Eq.~\eqref{eq:master} now gets generalized to
\begin{align}
\label{eq:master2}
\int d^3{x} \, & \langle \Delta\zeta({\vec x})\rangle_J  \, {\delta\Gamma[{\bar\zeta}, \zeta_b =0 ]\over \delta {\bar\zeta}(\vec x)} \\
& + \langle \Delta\zeta({\vec x})\rangle_0 \, {\delta\Gamma[{\bar\zeta}, \zeta_b =0 ]\over \delta {\bar\zeta_b}(\vec x)}
=0 \; , \nonumber
\end{align}
which is just the statement  that $\Gamma [\bar \zeta, \zeta_b =0 ]$ is invariant under a simultaneous transformation on $\bar \zeta$ and $\zeta_b$:
\be \label{Gamma symmetry}
\Gamma[\bar \zeta, \zeta_b=0] = \Gamma[   \bar \zeta + \langle \Delta\zeta \rangle_J , \zeta_b = \langle \Delta \zeta \rangle_0] \; .
\ee

From the viewpoint of our mathematical manipulations so far, $\bar \zeta$ and $\zeta_b $ are independent variables. However, the generating functional provides a non-trivial relation between them---the expected one: $\zeta_b(\vec x)$ is the configuration $\bar \zeta(\vec x)$ reduces to in the absence of external sources:
\be
\bar \zeta(\vec x)\big|_{J =0}  = {\delta W}/{\delta J(x)}\big|_{J=0} = \langle \zeta(\vec x) \rangle_{J=0} = \zeta_b (\vec x) \; .
\ee
Therefore, if we restrict the effective action $\Gamma$ to configurations that are localized perturbations---which we can turn on via localized sources---of some background field configuration, we can identify $ \zeta_b(\vec x)$ with such a  background configuration, which can be inferred from the behavior of $\bar \zeta(\vec x)$ at spatial infinity
\footnote{Notice that for this inference to be possible, it is crucial that the $\zeta_b(\vec x)$ backgrounds we are interested in are not generic functions of $\vec x$: they are characterized by a finite number of independent parameters---$\lambda$ and $\vec b$ in our case---, since they are generated by the action of a finite-dimensional symmetry group on the $\zeta_b=0$  configuration.}.
In such a case the effective action just becomes a functional of $\bar \zeta$, 
\be
\Gamma[\bar \zeta ] \equiv \Gamma[\bar \zeta ;  \zeta_b = \bar \zeta(\infty) ] \; ,
\ee
and now Eq.~\eqref{Gamma symmetry} reduces to the expected symmetry statement:
\be
\Gamma[\bar \zeta] = \Gamma[\bar \zeta + \langle \zeta \rangle_J] \; ,
\ee
where it is understood that $\bar \zeta (\vec  x)$ vanishes at spatial infinity (and, as a consequence, so does $J(\vec x)$).

To summarize: in sect.~\ref{formalism}, all statements based on symmetry  prior to Eq.~\eqref{sym} need qualifying, but the final symmetry statement on $\Gamma[\bar \zeta]$ is correct. Note that the same qualifications---and the same conclusion---apply to the case of spontaneously broken global symmetries.

%%%%%%%%%%%%%%%%%%%%%%%%%%%%
%%%%%%%%%%%%%%%%%%%%%%%%%%%%
\section{Ambiguity cancellation}\label{ambiguity}

Suppose we extrapolate our 1-PI vertices $\Gamma^{(n+1)}$, $\Gamma^{(n)}$ out of the momentum-conserving surface, by making them functions  
respectively of $n+1$ and $n$ {\em unconstrained} momenta. Since Eq.~\eqref{eq:mess} involves first derivatives on the l.h.s.~and second derivatives on the r.h.s, as well as momentum-conserving delta functions, we only need to consider displacements from the momentum-conserving surface that are of first order for the l.h.s., and up to second order for the r.h.s.:
\begin{align}
\Gamma^{(n+1)} & \to \Gamma^{(n+1)} + \gamma^{(n+1)} \equiv \Gamma^{(n+1)} + \vec u_{n+1} \cdot \vec g_{n+1} \label{extrapolate n+1}\\
\Gamma^{(n)} & \to \Gamma^{(n)} + \gamma^{(n)} \equiv \Gamma^{(n)} + \vec u_{n} \cdot \vec g_{n} + u_n^i u_n^j \, g^{ij}_n \; , \label{extrapolate n}
\end{align}
where the $\vec u \, $'s are the relevant total momenta,
\be
\vec u_{n+1} = \vec q + \sum_b \vec k_b \; , \qquad \vec u_{n} = \sum_b \vec k_b  \; ,
\ee
and the $g_{n+1}$, $g_n$ vectors and tensors are generic functions of all the $n+1$ and $n$ momenta involved.
If we now expand the derivatives of Eq.~\eqref{eq:mess} like we did in sect.~\ref{sec:special}, we end up with the following additional contributions. 

For the left hand side:
\be
2 \vec\nabla_{\vec q \to 0} \delta(\vec u_{n+1}) \, \gamma^{(n+1)} \Big|_{\vec q \to 0} +2  \delta(\vec u_{n}) \,  \vec g_{n+1} \Big|_{\vec q \to 0}  \; .
\ee

For the right hand side:
\begin{align}
2 \vec\nabla_{\vec q \to 0} & \delta(\vec u_{n+1})  \big(3 - {\cal D}_n \big) \gamma^{(n)} \\
&  +  \delta(\vec u_n) \Big[ 2(2- {\cal D}_n) \vec g_n + g^{ij}_n \sum_b \vec {\cal S}_b (u_n^i u_n^j) \Big]  \nonumber \; ,
\end{align}
where we have used the same identities for derivatives of the delta function as in sect.~\ref{sec:special}. We have also used the vanishing of $L^{ij} \gamma^{(n)} $ due to rotational invariance as well as  an analogous property of $\vec g_n$
\footnote{By rotational invariance, the vector nature of $\vec g_n$ can only come from its arguments---the momenta. We can thus write $\vec g_n = \sum_a \vec k _a \, g_a$, where the $g_a$ are scalar functions of the momenta, and use $L^{ij} g_a = 0$.}. 
It is a matter of simple algebra to show that
\be
\sum_b  {\cal S}_b^k (u_n^i u_n^j) = 2 \Big( \delta^{ij} u _n^k - \delta^{ik} u_n^j - \delta^{jk} u^i_n \Big) \; ,
\ee
which thus vanishes when multiplied by the delta function above. 

We are thus left with these new contributions to our special-conformal consistency relation:
\begin{align}
\mbox{lhs:} &  \quad 2 \vec\nabla_{\vec q \to 0} \delta(\vec u_{n+1}) \, \gamma^{(n+1)} \big|_{\vec q \to 0}+2  \delta(\vec u_{n}) \,\vec g_{n+1} \Big|_{\vec q \to 0}  \\
\mbox{rhs:} & \quad 2  \vec\nabla_{\vec q \to 0} \delta(\vec u_{n+1}) \big(3 - {\cal D}_n \big) \gamma^{(n)} + 2 \delta(\vec u_n) \, (2- {\cal D}_n) \vec g_n
\end{align}

The terms involving the derivative of the delta function cancel between l.h.s.~and r.h.s.~{\em only if} we assume that $\gamma^{(n+1)}$ and $\gamma^{(n)}$ obey the dilation consistency relation, that is, if we make sure that the extrapolations \eqref{extrapolate n+1} and \eqref{extrapolate n} preserve the dilation consistency relation, now between 1PI vertices evaluated at momenta that do not add up to zero. This is an in-principle non-trivial check one should perform on the candidate ``off-shell'' 1PI vertices. Once they pass it, one is then left with the terms involving the non-differentiated delta functions, which also vanish, for exactly the same reason: given the relation between the $\gamma$'s and the $\vec g \,$'s (see eqs.~\eqref{extrapolate n+1}, \eqref{extrapolate n}), the $\vec g$'s obey the dilation consistency relation
\be
\vec g_{n+1} \Big|_{\vec q \to 0} = (2- {\cal D}_n) \vec g_n \; 
\ee
as long as the $\gamma$'s obey the usual one.

\end{document}